\documentclass[12pt]{iopart}
\usepackage{iopams}
\usepackage {graphicx}
\begin{document}
\title{The shape of the orbit in FLRW spacetimes}

\author{David Garfinkle}

\address{Dept. of Physics, Oakland University, Rochester, MI 48309, USA}
\address{Michigan Center for Theoretical Physics, Randall Laboratory of Physics, University of Michigan, Ann Arbor, MI 48109-1120, USA}

\ead{garfinkl@oakland.edu}

\author{Lawrence R. Mead}

\address{Dept. of Physics and Astronomy, University of Southern Mississippi, Hattiesburg, MS 39406, USA}

\ead{lawrence.mead@usm.edu}

\author{H. I. Ringermacher}
\address{Dept. of Physics and Astronomy, University of Southern Mississippi, Hattiesburg, MS 39406, USA}

\ead{ringerha@gmail.com}

\begin{abstract}

The shape of the orbit of a free particle is examined in FLRW cosmologies.  The spatial projection of the orbit is time-independent and has a simple geometric description. We relate this description to the expression in terms of standard FLRW coordinates.

\end{abstract}

\maketitle

The spacetime of Friedmann-Lema\^itre-Robertson-Walker (FLRW) cosmology consists of a constant curvature space with a time dependent scale factor.  An object in free fall, or a light ray, travels along a spacetime geodesic, which traces out a path in the constant curvature 3-space.  It is a remarkable fact that this path is also a geodesic of the underlying constant curvature 3-space, and is therefore completely independent of the behavior of the scale factor.  Due to the symmetries of FLRW, there are a variety of geometric and coordinate based techniques that can be used to study geodesics.  The standard textbook approach\cite{Weinberg,MTW,Schutz,Wald,Ryden,Carroll} actually uses a combination of techniques: standard FLRW coordinates are used, but one makes use of homogeneity to demand that the geodesic passes through the point $r=0$ and is thus a radial geodesic.  This is convenient because the geodesics we are most interested in are those that reach our position, and we adopt for ourselves the coordinate priveleged position $r=0$.  Nonetheless, it is sometimes of interest to consider non-radial geodesics in FLRW, for example in the Sunyaev-Zeldovich effect\cite{SZ} where cosmic background photons inverse Compton scatter off the hot gas in a galaxy cluster; or more generally when treating the cosmological Boltzmann equation (see {\it e.g.} \cite{Dodelson}).  In this Note we will look at FLRW geodesics in general ({\it i.e.} not necessarily radial) form, using two different approaches: (1) a geometric approach and (2) a coordinate based approach.  We will present and compare these two approaches.

The metric $g_{ab}$ of the FLRW cosmology can be written as 
\begin{equation}
{g_{ab}} = {u_a}{u_b} + {a^2} {{\bar h}_{ab}}
\end{equation} 
Here ${u^a}={{(\partial /\partial t)}^a}$ is the four-velocity of the FLRW fluid, $a(t)$ is the scale factor, and 
${\bar h}_{ab}$ is the metric of a unit constant curvature space.  Note that this constant curvature space can be flat, positively curved (3-sphere), or negatively curved (hyperbolic space).
Let $k^a$ be the four-velocity of a timelike or null geodesics ({\it i.e.} an object in free-fall, or a light ray).  Then ${k^a}{k_a}=-\kappa$ where $\kappa =1 $ for material objects and $\kappa =0$ for light rays.  $k^a$ takes the form 
\begin{equation}
{k^a} = \alpha {u^a} + \beta {v^a}
\end{equation}
where $v^a$ is a unit vector in the metric ${\bar h}_{ab}$.  From the geodesic equation and the standard result ${\nabla _a}{u_b} = a {\dot a} {{\bar h}_{ab}} $ we find
\begin{equation}
{k^a}{\nabla _a} ({\alpha ^2} - \kappa ) = - {\frac 2 a} ({\alpha ^2} - \kappa ) {k^a}{\nabla _a} a
\; \; \; .
\end{equation}
It then follows that 
\begin{equation}
\alpha = {\sqrt {\kappa + {c_0 ^2}/{a^2}}} \; \; \; , \; \; \; \beta = {c_0}/{a^2}
\label{momentum}
\end{equation}
for some constant $c_0$.  (As shown in \cite {Carroll} eqn. (\ref{momentum}) can also be derived by first showing that ${a^4}{{\bar h}_{ab}}$ is a Killing tensor).  Now let $\xi ^a$ be a Killing vector of ${\bar h}_{ab}$.  Then it follows that $\xi ^a$ is also a Killing vector of the FLRW spacetime.  Thus there is a constant $c_1$ such that
\begin{equation}
{c_1}={g_{ab}}{k^a}{\xi^b}= \left ( {u_a}{u_b} + {a^2} {{\bar h}_{ab}} \right )
( \alpha {u^a} + \beta {v^a} ) {\xi ^b} = {c_0} {{\bar h}_{ab}} {v^a}{\xi ^b}
\end{equation} 
Thus $v^a$, the tangent vector to the spatial projection of the orbit, is a unit vector in 
the underlying unit constant curvature space whose inner product with every Killing vector of that space is a constant.  It then follows that the orbit is a geodesic of that space.  

We now describe the orbits for each possible curvature of space: for flat space, the orbit is a straight line.  For positive curvature, the orbit is a great circle of the 3-sphere.  That is, realizing the three sphere as the surface ${w^2} + {x^2} + {y^2} + {z^2} = 1$ in a flat 4-dimensional Euclidean space, the great circle is the intersection of this surface with a plane through the origin.  The analogous result holds for hyperbolic space.  Realizing this space as the surface (unit hyperboloid) $-{t^2} + {x^2} + {y^2} + {z^2} = -1$ in Minkowski spacetime, the orbit is the intersection of this surface with a plane through the origin.

We now consider the coordinate description of the orbit.  It is both convenient and usual in treating 
FLRW spacetimes to use a single expression, subsuming all three possibilities for the curvature, to describe the spacetime.  The line element is given by the expression
\begin{equation}
d {s^2} = - d {t^2} + {a^2} (t) \left [ {\frac {d {r^2}} {1 - k {r^2}}} + {r^2} ( d {\theta ^2} + {\sin ^2} \theta d {\varphi ^2} ) \right ] \; \; \; ,
\label{metric}
\end{equation}
where $k=0$ (flat space), $k=1$ (3-sphere), or $k=-1$ (hyperbolic space).  The geodesic equation is
\begin{equation}
{\frac {{d^2}{x^\mu}} {d{\lambda ^2}}} + {\Gamma ^\mu _{\alpha \beta}} {\frac {d {x^\alpha}} {d \lambda}}
{\frac {d {x^\beta}} {d \lambda}} = 0 \; \; \; ,
\end{equation}
where $\lambda $ is an affine parameter.  Without loss of generality, we specialize to orbits in the $\theta = \pi/2$ plane.  The geodesic equation then yields
\begin{eqnarray}
{\frac {{d^2}t} {d {\lambda ^2}}} + a {\dot a} \left [ {\frac 1 {1- k{r^2}}}  {{\left ( {\frac {dr} {d \lambda}}\right ) }^2} + {r^2} {{\left ( {\frac {d\varphi} {d \lambda}}\right ) }^2}   \right ] = 0 \; \; \; ,
\label{tgeo}
\\
{\frac {{d^2}r} {d {\lambda ^2}}} + {\frac {kr} {1 - k {r^2}}} {{\left ( {\frac {dr} {d \lambda}}\right ) }^2} - r (1 - k {r^2}) {{\left ( {\frac {d\varphi} {d \lambda}}\right ) }^2} + 2 {\frac {\dot a} a} 
{\frac {dr} {d\lambda}}{\frac {dt} {d\lambda}} = 0 \; \; \; ,
\label{rgeo}
\\
{\frac {{d^2}\varphi} {d {\lambda ^2}}} + 2 {\frac {\dot a} a} {\frac {d \varphi} {d \lambda}} {\frac {dt} {d\lambda}} + {\frac 2 r} {\frac {d \varphi} {d \lambda}} {\frac {dr} {d\lambda}} = 0 \; \; \; .
\label{phigeo}
\end{eqnarray}
Defining $d/d\ell = {a^2} d/d\lambda$ along the geodesics, one obtains from eqns. (\ref{rgeo}-\ref{phigeo}) 
\begin{eqnarray}
{\frac {{d^2}r} {d {\ell ^2}}}  + {\frac {kr} {1 - k {r^2}}} {{\left ( {\frac {dr} {d \ell}}\right ) }^2} - r (1 - k {r^2}) {{\left ( {\frac {d\varphi} {d \ell}}\right ) }^2} = 0 \; \; \; ,
\label{rgeo2}
\\
{\frac {{d^2}\varphi} {d {\ell ^2}}} + {\frac 2 r} {\frac {d \varphi} {d \ell}} {\frac {dr} {d\ell}} = 0 \; \; \; ,
\label{phigeo2}
\end{eqnarray}
thus demonstrating that the shape of the orbit is independent of the scale factor.  And, in fact, eqns. (\ref{rgeo2}-\ref{phigeo2}) are precisely the geodesic equations for the metric ${\bar h}_{ab}$ with $\ell$ as the length in that metric.  One solution of eqn. (\ref{phigeo2}) is 
$d\varphi /d\ell =0$ which corresponds to radial geodesics.  In this case, the first integral of eqn. (\ref{rgeo2}) leads to $d\ell = dr/\sqrt{1-k{r^2}}$. This standard result, which can also be read off directly from eqn. (\ref{metric}), in turn leads to the standard result that 
\begin{equation}
r=\ell \, (k=0) \; \; \; , \; r= \sin \ell \, (k=1) \; \; \; , \; r = \sinh \ell \, (k=-1) \; \; \; . 
\label{radial}
\end{equation}
If $d\varphi /d\ell \ne 0$ (non-radial geodesics) then using eqn. (\ref{phigeo2}) and the chain rule in eqn. (\ref{rgeo2}) we obtain
\begin{equation}
{\frac {{d^2}r} {d {\varphi ^2}}} - {\frac {2 - 3 k {r^2}} {r(1-k{r^2})}}  {{\left ( {\frac {dr} {d \varphi}}\right ) }^2} - r (1 - k {r^2}) = 0 \; \; \; .
\label{shape}
\end{equation}
Defining $u \equiv {r^{-2}} - k$, eqn. (\ref{shape}) becomes
\begin{equation}
{\frac {{d^2}u} {d{\varphi ^2}}} - {\frac 1 {2u}} {{\left ( {\frac {du} {d \varphi}}\right ) }^2}
+ 2 u = 0 \; \; \; ,
\label{shape2} 
\end{equation}
from which it follows that
\begin{equation}
{\frac d {d \varphi}} \left [ {u^{-1}} {{\left ( {\frac {du} {d \varphi}}\right ) }^2} + 4 u \right ] = 0 \; \; \; .
\label{shape3}
\end{equation}
Therefore there is an integration constant $c_1$ for which 
\begin{equation}
{u^{-1}} {{\left ( {\frac {du} {d \varphi}}\right ) }^2} + 4 u  = {c_1} \; \; \; .
\label{shapeintegral}
\end{equation}
Note that since $r \le 1$ for $k=1$, it follows that $u \ge 0$ for any value of $k$, and thus using eqn. (\ref{shapeintegral}) it follows that ${c_1} \ge 0$.  Now defining $w \equiv u - {c_1}/8$ we find that eqn. (\ref{shapeintegral}) is equivalent to
\begin{equation}
{{\left ( {\frac {dw} {d \varphi}}\right ) }^2} + 4 {w^2} = {\frac {c_1 ^2} {16}} \; \; \; .
\label{shapeintegral2}
\end{equation}
Now differentiating eqn. (\ref{shapeintegral2}) with respect to $\varphi$ we obtain
\begin{equation}
{\frac {{d^2} w} {d {\varphi ^2}}} + 4 w = 0 \; \; \; .
\label{shape4}
\end{equation}
The general solution to eqn. (\ref{shape4}) is 
\begin{equation}
w = A \cos (2 (\varphi + \beta ))
\label{wsoln}
\end{equation} 
where $A$ is a nonnegative constant, and $\beta$ is a constant.  Now using eqn. (\ref{wsoln}) in eqn. (\ref{shapeintegral2}) we find that the integration constant is given by ${c_1} = 8 A$ (where we have also used the fact that $c_1$ is nonnegative).  It then follows from the definition of $w$ that
\begin{equation}
u = 2 A {\cos ^2} (\varphi + \beta) \; \; \; ,
\label{usoln}
\end{equation}  
And thus from the definition of $u$ that 
\begin{equation}
r = {\frac 1 {\sqrt {k + 2 A {\cos ^2} (\varphi + \beta)}}} \; \; \; .
\label{rsoln}
\end{equation}

We now compare eqn. (\ref{rsoln}) to the results of the geometric approach for the three cases: $k=0,1,-1$.  For $k=0$ the space is flat and $(r,\varphi)$ are plane polar coordinates.  However, in this case eqn. (\ref{rsoln}) is easily seen to be the equation for a straight line in plane polar coordinates, so the geometric result agrees with the coordinate result.  

\begin{figure}
\centering
\includegraphics[width=0.75\textwidth]{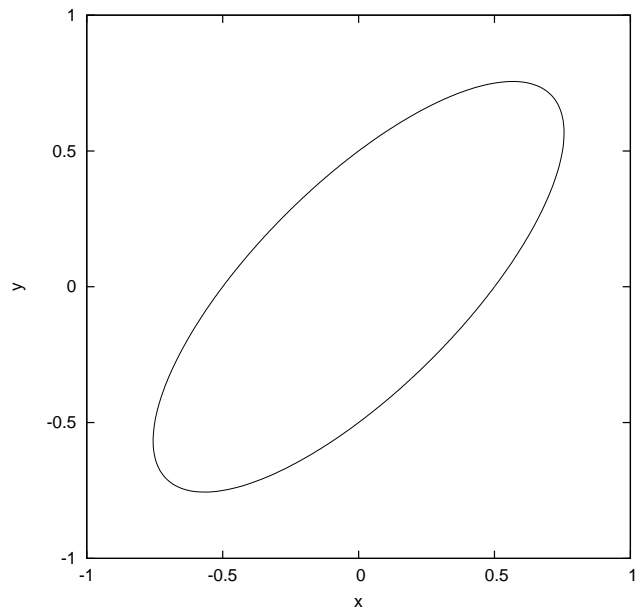}
\caption{The orbit in the $k=1$ case using coordinates $(x=r\cos \varphi, y=r\sin \varphi)$.  Here $A=3$ and $\beta=\pi/4$.}
\label{fig:1}
\end{figure}

For $k=1$, the geometric result is that the orbit is a circle.  However, eqn. (\ref{rsoln}) does not look like the equation for a circle.  In fact, if $(r,\varphi)$ were plane polar coordinates, then eqn. (\ref{rsoln}) would be the equation of an ellipse (see Fig. \ref{fig:1}).  The coordinates $(w,x,y,z)$ for the embedding of the 3-sphere in 4 dimensional flat space are related to the FLRW coordinates $(r,\varphi)$ by
\begin{equation}
(w,x,y,z) = ({\sqrt {1-{r^2}}},r \cos \varphi, r \sin \varphi, 0)
\label{3sph}
\end{equation}
where we have also used the condition $\theta = \pi/2$. The equation for a plane through the origin is $w = {c_2} x + {c_3} y $ for some constants $c_2$ and $c_3$, which becomes using eqn. (\ref{3sph})
\begin{equation}
{\sqrt {1 - {r^2}}} = B r \cos (\varphi + \beta )
\end{equation}
for some constants $B$ and $\beta$.  Solving for $r$ we obtain
\begin{equation}
r = {\frac 1 {\sqrt {1 + {B^2} {\cos ^2} (\varphi + \beta)}}} \; \; \; .
\label{rsoln2}
\end{equation}
which agrees with eqn. (\ref{rsoln}).

\begin{figure}
\centering
\includegraphics[width=0.75\textwidth]{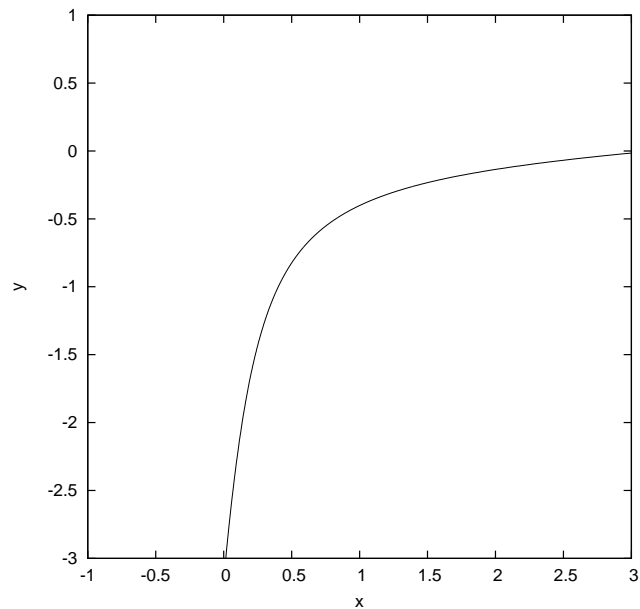}
\caption{The orbit in the $k=-1$ case using coordinates $(x=r\cos \varphi, y=r\sin \varphi)$.  Here $A=1.1$ and $\beta=\pi/4$.}
\label{fig:2}
\end{figure}

Corresponding results hold for $k=-1$ (see Fig. \ref{fig:2}). We have 
\begin{equation}
(t,x,y,z) = ({\sqrt {1+{r^2}}},r \cos \varphi, r \sin \varphi, 0)
\label{3hyp}
\end{equation}
where we have also used the condition $\theta = \pi/2$. The equation for a plane through the origin is $t = {c_2} x + {c_3} y $ for some constants $c_2$ and $c_3$, which becomes using eqn. (\ref{3hyp})
\begin{equation}
{\sqrt {1 + {r^2}}} = B r \cos (\varphi + \beta )
\end{equation}
for some constants $B$ and $\beta$.  Solving for $r$ we obtain
\begin{equation}
r = {\frac 1 {\sqrt {-1 + {B^2} {\cos ^2} (\varphi + \beta)}}} \; \; \; .
\label{rsoln3}
\end{equation}
which agrees with eqn. (\ref{rsoln}).

Thus we see that for a given value of $k$ all the FLRW orbits are the same geometrically, and are therefore equivalent to orbits of the standard radial geodesics of eqn. (\ref{radial}).  However, these orbits {\emph {look}} very different for non-radial geodesics expressed in standard FLRW coordinates.  This is completely analogous to the well known phenomenon (familiar to airplane travelers) that though all great circles of the Earth are geometrically equivalent to the equator, they look very different when expressed in terms of longitude and latitude.   

\section*{Acknowledgements}
This work was supported by NSF grants PHY-1505565 and PHY-1806219 to Oakland University.

\end{document}